\documentclass[twocolumn,twoside,aps,superscriptaddress,prl,showpacs]{revtex4}

\usepackage{amsmath,amssymb,graphicx}

\newcommand{\Ha}{{\mathbf H}}
\newcommand{\M}{{\mathbf M}}

\newcommand{\e}{{\mathbf e}}
\newcommand{\om}{\omega}
\newcommand{\pa}{\partial}
\newcommand{\na}{\nabla}
\newcommand{\al}{\alpha}
\newcommand{\N}{{\mathbf N}}
\newcommand{\m}{{\mathbf m}}

\begin{document}

\title{Ferrofluids as thermal ratchets}

\author{Andreas Engel}
\email[]{andreas.engel@physik.uni-magdeburg.de}
\affiliation{Universit\"at Magdeburg, Institut f\"ur Theoretische
Physik,
  PSF 4120, 39106 Magdeburg, Germany}
\author{Hanns Walter M\"uller}
\affiliation{Max-Planck-Institut f\"ur Polymerforschung, 55128
Mainz, Germany}
\author{Peter Reimann}
\affiliation{Fakult\"at f\"ur Physik, Universit\"at Bielefeld,
        PSF 100131, 33501 Bielefeld, Germany}
\author{Achim Jung}
\affiliation{Technische Physik, Universit\"{a}t des Saarlandes, 
  PSF 1551150, 66041 Saarbr\"{u}cken, Germany}

\date{\today}

\begin{abstract}
Colloidal suspensions of ferromagnetic nano-particles, so-called
ferrofluids, are shown to be suitable systems to demonstrate and
investigate thermal ratchet behavior: By rectifying thermal
fluctuations, angular momentum is transferred to a resting ferrofluid
from an oscillating magnetic field without net rotating component. Via
viscous coupling the noise driven rotation of the microscopic
ferromagnetic grains is transmitted to the carrier liquid to yield 
a macroscopic torque. For a simple setup we analyze the rotation of the 
ferrofluid theoretically and show that the results are compatible with
the outcome of a simple demonstration experiment.   
\end{abstract}


\maketitle

To extract directed motion from random fluctuations is a  
problem at the heart of statistical mechanics with long history 
\cite{Maxwell,Smo,Fey,HaRe}. On the one hand it is crucial for basing
the second law of thermodynamics on statistical reasoning and for
depriving various Maxwell's demon type devices of their mystique.
On the other hand it is closely linked to practical issues like
the efficiency of heat engines. The problem has gained renewed
attention \cite{HaBa,Ast,Reirev} due to its possible relevance for
biological transport \cite{Mag,JAP} and the prospects of nano-technology
\cite{RSAP,OuBo,Lin}. 

A particularly clear example of a so-called {\em ratchet}
is provided by an over-damped particle in a time
dependent on-off spatially sawtooth potential subject to some
random noise \cite{AjPr}. In the stationary (on or off) state,
thermodynamic equilibrium prevails and detailed balance prohibits
directed transport. Complementary, with the potential
switched on and off regularly diffusion and relaxation combine to
yield a non-zero average particle drift. Other means may be used to
drive the system away from equilibrium. In any case a noise driven
drift is to be expected generically under non-equilibrium conditions
where, however, the specification of the value and even the sign of
this drift can be rather subtle \cite{Reirev}.

In the present letter we show that colloidal suspensions of ferromagnetic
nano-particles, so-called ferrofluids \cite{Ros}, are ideal systems to test
theoretical predictions on fluctuation driven transport experimentally. This
is due to three main reasons: First, the small size of the ferromagnetic
grains ($\sim 10$nm) implies that their dynamics is strongly
influenced by thermal fluctuations \cite{Sh}. Second, spatially
periodic time-dependent potentials for the orientation of the
particles can be easily realized by external magnetic fields. Third,
directed orientational transport manifests itself as systematic
rotation of the ferromagnetic particles which can be easily detected
from the resulting macroscopic torque on the carrier liquid. We show
that our theoretical consideration are in agreement with a simple
demonstration experiment. 

Ferrofluids combine the hydrodynamic properties of
Newtonian fluids with the magnetic behavior of superparamagnets
\cite{Ros}. Many of their fascinating properties stem from the
viscous coupling between the rotation of the magnetic grains and the
vorticity of the hydrodynamic flow. 
A direct way to set this coupling into action is with a {\em rotating}
external magnetic field as used, e.g., to spin up a ferromagnetic
drop floating under hydraulic zero-gravity \cite{BCP,MEL}.
Complementary, a rotational ferrofluid flow exposed to a static
magnetic field exhibits an enhanced shear viscosity
\cite{McTague}. Various unusual effects such as ``negative''
rotational viscosity \cite{ShMo}, magneto-vortical resonance
\cite{GHBCP}, and anomalously enhanced ac-response due to
coherent particle rotation \cite{HWM} rely on the exchange of
angular momentum between {\em rotating} particles and an {\em
  oscillating} magnetic field. In these cases the 
involved non-zero flow vorticity is crucial for explicitly
breaking the symmetry between clockwise and counter-clockwise
particle rotation.

In contrast to these situations we demonstrate the transfer of 
angular momentum from an {\em oscillating} magnetic field to a
ferrofluid {\em at rest}. A 
crucial ingredient of the operating mechanism is a sufficient
impact of thermal noise due to random collisions between the
ferromagnetic grains and the molecules of the carrier liquid.

We consider a ferrofluid in a horizontal, spatially homogeneous
magnetic field $\Ha$ composed of a constant part $H_x$ parallel to the
$x$-axis and an oscillatory part $H_y f(t)$ along the
$y$-direction. The precise form of the time modulation $f(t)$ will be
fixed below. As common for diluted ferrofluids \cite{Ros,Sh} we model
the ferromagnetic particles as non-interacting spherical equal sized
dipoles of volume $V$ and magnetic moment $\m$ \cite{com1}. The
external magnetic field gives rise to a potential $U=-\m\cdot\Ha$
for the direction of the magnetic moment. 
This direction in turn is assumed to be tightly co-aligned with
the orientation of the magnetic particle. A reorientation of the
magnetic moment hence requires a rotation of the particle against the
viscosity $\eta$ of the carrier liquid characterized by the Brownian
relaxation time $\tau_B=3 \eta V/k_B T$ with $T$ denoting the
temperature of the liquid. 
We use the dimensionless Langevin parameters 
$\alpha_{x,y}=mH_{x,y}/k_B T$ to measure the magnetic field strengths
and scale time by twice the Brownian relaxation time.
As time dependence in the ac part of the field we choose
\begin{equation}\label{foft}
  f(t)=\cos(\om t)+a\sin(2\om t +\beta).
\end{equation}
The parameters $\alpha_x, \, \alpha_y, \, a, \, \omega$, and $\beta$ are
easily controlled in experiments. 

\begin{figure}
\includegraphics[width=\columnwidth]{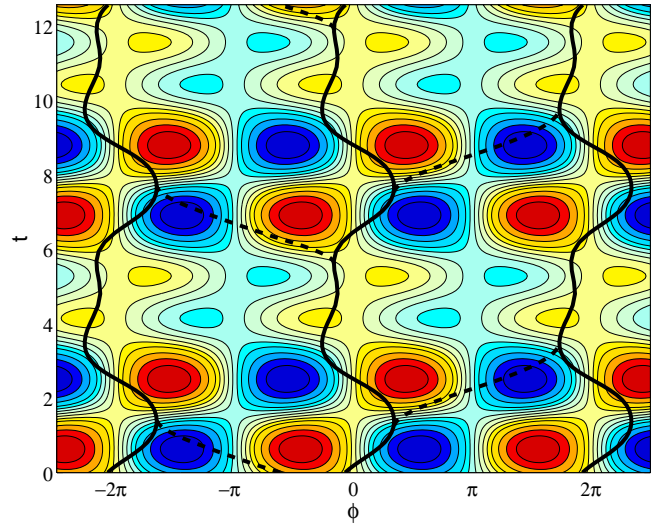}%
\caption{\label{fig1} Space-time plot of the potential
$U(\theta=\pi/2,\phi,t)$ from Eq.(\ref{pot}) for 
$\alpha_x=0.3, \,\alpha_y=1,\, a=1, \, \omega=1$, and $\beta=0$. 
Blue and red regions correspond to small and large values of $U$
respectively. In the long-time limit, the deterministic dynamics
approaches $\theta(t) =\pi/2$ and a 
periodic $\phi(t)$, represented by either of the full black lines.
In the presence of thermal noise, transitions between these
deterministic solutions become possible schematically indicated by
the dashed lines. The spatial asymmetry and temporal anharmonicity of
the potential results in slightly different rates for 
noise induced increments and decrements of $\phi$ respectively. As
a result a noise driven rotation of the particles arises.}
\end{figure}

Parameterizing the orientation of a magnetic dipole by two 
angles, $\e=\m/m=(\sin\theta\cos\phi,\sin\theta\sin\phi,\cos\theta)$  
the potential $U$ generated by the above specified magnetic field 
takes the form
\begin{equation}\label{pot}
  U(\theta,\phi,t)= -\sin\theta(\alpha_x \cos\phi+ \alpha_y f(t)\sin\phi).
\end{equation}
Neglecting thermal fluctuations, the particle orientation
is governed by an overdamped relaxation dynamics in the
force field $- \na U$. Accordingly, the orientation $\e (t)$ 
tends to align with the magnetic field. Since the field is
restricted to the $x$-$y$-plane, $\theta (t)\to \pi/2$ for $t\to\infty$.
Within the plane, the angle between the field and
the $x$-axis is always less than $\pi/2$, implying the same for
$\phi(t)$ when $t\to\infty$, see Fig.\ref{fig1}, i.e. no average
rotation or torque can arise. 

Things change in the presence of thermal fluctuations 
causing stochastic transitions between the deterministic solutions. As
elucidated qualitatively with Fig.\ref{fig1} the dynamical asymmetry
induced by $\alpha_x\neq 0$ and $a\neq 0$ gives rise to slightly different
probabilities for a $2\pi$-phase-slip in the ``forward-$\phi$
direction'' and ``backward-$\phi$ direction'' respectively. This in
turn results in an average rotation of the particle, a manifestation
of the ratchet effect \cite{HaBa,Ast,PaWe} for the orientational
motion of the ferromagnetic grains. Similar to situations with
additive driving \cite{ChMi,SeMa,GoHa,Flach} the spatial symmetry
of our potential $U(\theta,\phi,t)$ requires a sufficiently complex 
time modulation function $f(t)$ for noise induced transport to occur. 

A more quantitative analysis of the effect can be made on the basis of
the Fokker-Planck equation for the probability distribution
$P(\theta,\phi,t)$ of the particle orientation. It is given by 
(see e.g. \cite{CKW})
\begin{equation}\label{FPE}
  \pa_t P=\na(P\;\na\; U)+\na^2 P, 
\end{equation}
where $\nabla$ denotes the angular part of the Nabla-operator in 
spherical coordinates.

We have solved this equation numerically by expanding 
$P(\theta,\phi,t)$ in spherical harmonics. From the solution we can
determine the average orientation 
$\langle \e \rangle = \int d\phi d\theta \sin\theta\,\e \,
P(\theta,\phi,t)$ and the average torque 
$\N_p=m \langle \e \rangle \times \Ha$ acting upon an individual
particle. For symmetry reasons only the $z$-component of the torque is
different from zero. Due to the viscous coupling between the
ferromagnetic grains and the carrier liquid the individual torques
combine to yield a macroscopic torque per fluid volume  
\begin{equation}\label{torque}
  {\bf N}=n \N_p= \mu_0 \, M_s \langle {\e} \rangle \times {\Ha}=
  \mu_0 \, {\M} \times {\Ha}.
\end{equation}
Here $n$ denotes the number density of the particles, 
$\M=M_s \langle{\bf e} \rangle$ the magnetization of the ferrofluid, 
$M_s=m n/\mu_0$ its saturation value and $\mu_0$ the permeability of free
space. A non-zero torque after averaging (\ref{torque}) over one
period of the external ac-driving is the macroscopic signature of a
directed orientational transport. 

An explicit expression for the torque can be obtained from an approximate
solution of the Fokker-Planck equation (\ref{FPE}) by adopting the effective 
field method \cite{MRS}. It gives rise to a relaxation equation \cite{MuLi} 
for the average orientation $\langle \e \rangle$ which for a fluid at rest
reduces to 
\begin{equation}\label{mdot}
\partial_t \langle e_i \rangle= - \sum_j\left[
   (1-\frac{ \langle e \rangle}{\al_{\text{eff}}})\delta_{ij}
   + (3\frac{ \langle e \rangle}{\al_{\text{eff}}}-1) 
     \frac{\langle e_i \rangle}{\langle e \rangle}
     \frac{\langle e_j \rangle}{\langle e \rangle}\right ] 
      (\al_{\text{eff}}-\al)_j
\end{equation}
where $\langle e \rangle$ denotes $|\langle \e \rangle|$ and the
so-called effective field $\al_{\text{eff}}(\langle e 
\rangle)$ is given by the inverse of the Langevin function 
$\langle e \rangle={\cal L}(\alpha)=\coth{\alpha} - 1/\alpha$. 
The numerical integration of 
Eq.(\ref{mdot}) yields rather accurate approximations for the
orientation $\langle e_i \rangle(t)$, which deviate from the numerical
solution of the Fokker-Planck equation (\ref{FPE}) by
less than a few percent. Nevertheless the final values for the
time averaged magnetic torque (\ref{torque}) differ by a factor
between 2 and 3. This is due to the fact that the time average
$\overline{N_z}$ is much smaller than the typical values 
of the time dependent component $N_z$ of the torque. 

For sufficiently weak magnetic field we may use ${\cal L}(\alpha)\simeq
\al/3$ which makes eq.(\ref{mdot}) linear in $\langle \e \rangle$. It
is then easily solved for a time dependence of the field
$\al$ as specified by eq.(\ref{foft}). The resulting time averaged
torque (\ref{torque}), however, vanishes identically. Exploring the
non-linear regime perturbatively by using 
${\cal L}(\alpha)\simeq\alpha/3-\alpha^3/45$ we find to leading order
for the time-averaged $z$-component of the torque 
\begin{equation}\label{torquean}
\overline{N_z} = \mu_0 \, \frac{M_s^2}{3 \chi} \; \alpha_y^3\,\alpha_x
  \,\frac{a}{30} \;\frac{\omega^2 \, (\omega \cos{\beta} + 2 \sin\beta)}
  {(1+ \omega^2)(4 + \omega^2)^2}.
\end{equation}
where $\chi$ denotes the magnetic susceptibility of the ferrofluid. 
This expression provides a useful approximate formula for the
nontrivial dependencies of the torque on the parameters of the
problem. In particular it shows that both the static 
magnetic field component $\alpha_x$ and the anharmonic time dependence
(\ref{foft}) of the oscillatory component are essential for directed
rotational transport to occur. The effective field method may be an
useful tool for the analysis of other ratchet systems as well \cite{EMR}.  

\begin{figure}
\includegraphics[width=\columnwidth]{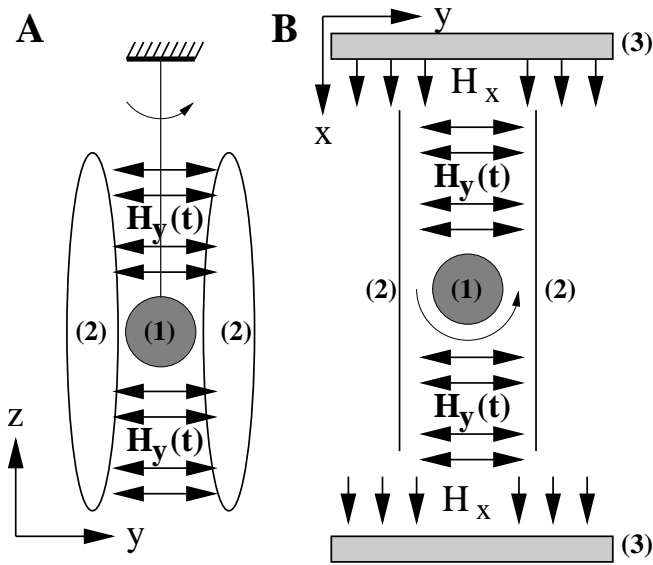}%
\caption{\label{fig2} Sketch of the experimental setup. ( A side view,
  B top view). A hollow plastic 
  sphere (1) is filled with ferrofluid and suspended on a thin Kevlar
  fiber. The time dependent magnetic field in $y$-direction is 
  generated with a pair of Helmholtz coils (2), the static field in
  $x$-direction stems from a commercial electro-magnet (3). Noise assisted
  transfer of angular momentum from the magnetic field to the
  ferrofluid manifests itself in a rotation of the sphere.}
\end{figure}

To verify our theoretical findings we have designed a simple
demonstration experiment as sketched in Fig.\ref{fig2}. 
The setup is a torsion balance similar to those used in string galvanometers. 
A hollow plastic sphere with inner diameter $16$ mm was filled with a
ferrofluid APG 933 (Ferro-Tec), with the following specifications:
density $\rho=1,100$~kg/${\rm m}^3$, susceptibility $\chi=1.09$,
saturation magnetization $M_s=18$~kA/m, dynamic viscosity $\eta=0.1$ Pas. 
The container was suspended on a Kevlar fiber with a length of $20$~cm
and $10 \, \mu$m diameter. The oscillatory magnetic field
was generated by a pair of Helmholtz coils of $100$ windings each via a 
computer generated signal of the form specified in Eq.(\ref{foft}).
The RMS-field strength in the center of the coil amounted to about
$H_y \simeq 2.1$~kA/m. The static field component $H_x\simeq 9$~kA/m was
generated by a commercial electro-magnet (Bruker). The applied frequency was
$\nu=200$~Hz. In the choice of the frequency and the ratio
between the amplitudes of the static and oscillating field we where
guided by our numerical solutions of the Fokker-Planck equation
(\ref{FPE}). The dimensionless parameters describing  the setup are
$\alpha_x=1.72,\, \alpha_y=0.4,\, a=1$. The dimensionless frequency 
$\om$ depends on the Brownian relaxation time $\tau_B$ which was used
as a fit parameter. 

The main qualitative result of the experiment is the unambiguous
demonstration of the proposed thermal ratchet effect: After switching
on the fields the ferrofluid sphere immediately starts to rotate. 
Switching off either the static field $\al_x$ or the modulation
amplitude $a$ the torque disappears. A reversal of the sign of either
$\al_x$ or $a$ causes a reversal of the sample's rotation sense. 

\begin{figure}
\includegraphics[width=\columnwidth]{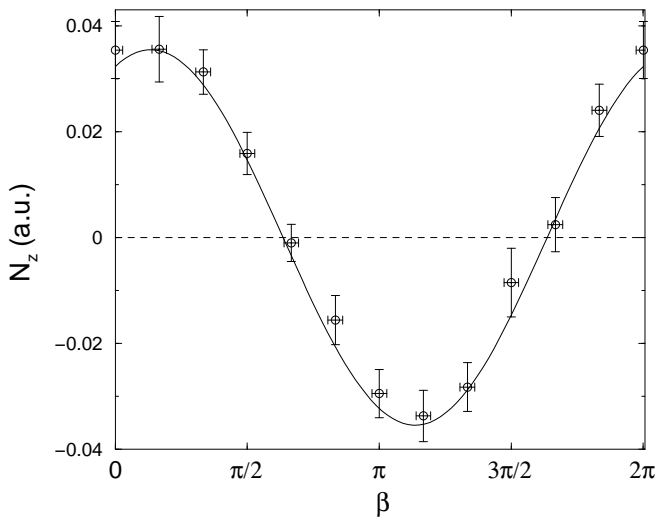}%
\caption{\label{fig3} Magnetic torque transferred to the ferrofluid as
  function of the phase angle $\beta$ for $\alpha_x=1.72,\,
  \alpha_y=0.4,\, a=1$. The symbols are experimental 
  results obtained with the setup shown in Fig.\ref{fig2}. The solid
  line is a fit to the approximate analytical expression (\ref{torquean}), 
  $N_z=A(\om \cos\beta+2\sin\beta)$ with the amplitude $A$ and the
  frequency $\om$ as fit parameters. The value $\om\simeq 4.41$ obtained
  translates into a fit for the Brownian relaxation time of 
  $\tau_B\simeq 1.8$ ms.}
\end{figure}

A more quantitative comparison between theory and experiment was difficult 
for the following reason: Our setup was designed to be as sensitive as 
possible with respect to the appearance or not of a rotation of the drop. 
The actual rotation speeds of the sphere then became usually so large
that the field had to be switched off after a short time in order not
to destroy the experiment. We hence resorted to the following
semi-quantitative procedure to determine the torque: Starting with the
sphere at rest and then letting act the magnetic field for a  
fixed amount of time (a few seconds), the resulting final rotation
speed of the sphere was used as a
measure for the torque. In this way we have experimentally recorded
the torque (in arbitrary units) as a function of the phase angle
$\beta$ in (\ref{foft}). The results are shown in fig.\ref{fig3}
together with a fit to our approximate theoretical result given  
in eq.(\ref{torquean}). 

Theory and experiment are compatible if the Brownian relaxation time
is chosen as $\tau_B\simeq 1.8$~ms. Using $\tau_B=3\eta V/k_B T$ this 
corresponds to a particle diameter of about 35 nm, which exceeds the
typical size by about a factor of 3. However, in real ferrofluids
size, shape, and magnetic moment of the ferromagnetic grains vary
quite significantly (see e.g. \cite{Ros}, page 41), a situation only
poorly described by our model involving a single relaxation
time. Moreover, since the transferred angular
momentum increases with the particle diameter \cite{EMR} the above
value for $\tau_B$ is likely to characterize the largest grains in the 
population rather than the average.  

In conclusion we have shown that by rectifying rotational Brownian motion
angular momentum can be transferred from an oscillating magnetic field
to a ferrofluid at rest. A unique feature ferrofluids offer in
comparison with other experimental realizations of ratchets is the
combined action of many individual nano-scale ratchets to yield a
{\it macroscopic} thermal noise induced transport effect. More precise  
experiments are needed to verify quantitatively our theoretical
predictions about the dependence of the torque on the magnetic field
strength and the frequency of the external driving. 

Many other investigations using ferrofluids as thermal ratchets 
are conceivable. A particularly interesting line is
linked with the occurrence of inversion points $\beta_i$ of the
rotation, see Fig.~\ref{fig3}. When keeping $\beta$ fixed to such an
inversion point and 
varying instead another parameter of the system the rotational speed
of the magnetic grains will generically exhibit a change of sign as a
function of this parameter as well. Since $\alpha_{x,y}$ or the
dimensionless $\omega$ (due to its dependence on $\tau_B$) depend on
the size of the magnetic grains in a polydisperse ferrofluid under
the same experimental conditions the larger and smaller particles may
rotate in opposite directions.

\acknowledgements Helpful discussions with K. Knorr, J. Albers,
P. Bl\"umler, I. Rehberg, and M. Raible are gratefully acknowleged. 
This work was supported by Deutsche Forschungsgemeinschaft under 
SFB 613, SFB 277, FOR 301 and the ESF-program STOCHDYN.

\end{document}